\documentclass[reprint,amsmath,amssymb,aps,prb,superscriptaddress]{revtex4-1}
\pdfoutput=1
\usepackage{hyperref,url}
\usepackage{amsmath}
\usepackage{amsfonts}
\usepackage[capitalise]{cleveref}
\usepackage{placeins}
\hypersetup{breaklinks,colorlinks=true,linkcolor=blue,citecolor=blue,urlcolor=blue,filecolor=blue}
\usepackage{graphicx}
\usepackage{dcolumn}

\newcommand{\EPT}{E_\text{PT2}}

\begin{document}

\author{Fionn D. Malone}
\thanks{These authors contributed equally}
\affiliation{Quantum Simulations Group, Lawrence Livermore National Laboratory, 7000 East Avenue, Livermore, CA, 94551 USA.}
\author{Anouar Benali}
\thanks{These authors contributed equally}
\affiliation{Computational Science Division, Argonne National Laboratory, Argonne, Illinois 60439 USA}
\author{Miguel A. Morales}
\affiliation{Quantum Simulations Group, Lawrence Livermore National Laboratory, 7000 East Avenue, Livermore, CA, 94551 USA.}
\author{Michel Caffarel}
\affiliation{Laboratoire de Chimie et Physique Quantiques, Université de Toulouse, CNRS, UPS,
France}
\author{P. R. C. Kent}
\affiliation{Center for Nanophase Materials Sciences and Computational Sciences and Engineering Division, Oak Ridge National Laboratory, Oak Ridge, Tennessee 37831, USA.}
\author{Luke Shulenburger}
\email{lshulen@sandia.gov}
\affiliation{HEDP Theory Department, Sandia National Laboratories, Albuquerque, New Mexico 87185 USA}

\title{Systematic Comparison and Cross-validation of Fixed-Node Diffusion Monte Carlo and Phaseless Auxiliary-Field Quantum Monte Carlo in Solids}

\begin{abstract}
Quantum Monte Carlo (QMC) methods are some of the most accurate methods for simulating correlated electronic systems. We investigate  the  compatibility,  strengths  and  weaknesses  of  two such  methods,  namely,  diffusion  Monte  Carlo  (DMC)
and  auxiliary-field  quantum  Monte  Carlo (AFQMC). The multi-determinant trial wave functions employed in both approaches are generated using
the configuration interaction using a perturbative selection  made  iteratively  (CIPSI)  technique. Complete basis set
full configuration interaction (CBS-FCI) energies estimated with CIPSI are used as a reference in this comparative study between DMC and AFQMC.
By focusing on a set of canonical finite size solid state systems, we show
that both QMC methods can be made to systematically converge towards the same energy once basis set effects and systematic biases
have been removed. AFQMC shows a much smaller dependence on the trial wavefunction than DMC while simultaneously exhibiting a much
larger basis set dependence. We outline some of the remaining challenges and opportunities for improving these approaches.
\end{abstract}
\maketitle

\emph{Introduction.}
The accurate first principles description of correlated materials is one of the grand challenges of chemistry, materials science
and physics\citep{Kent_strong_corr_2018}. Density functional theory\citep{dft1hk,dft2ks} (DFT) is the workhorse of these
communities, offering an often good enough accuracy relative to its computational cost. However, use of DFT in practice suffers
from a number of well known deficiencies including uncertainty in the choice of exchange-correlation
functional\citep{Medvedev_functional_2017} and with the treatment of strongly correlated materials. While no single approach is
likely to work well in every situation\citep{Carter_challenges_08}, complementary methods are desired that can be systematically
converged and applied to novel materials in a fully ab initio manner\citep{booth2013towards}. Here we focus on quantum Monte Carlo (QMC) methods that can
potentially achieve this goal.

Several different flavors of QMC exist. Generally, ground-state QMC methods use a direct wavefunction based approach to solving
the many-electron Schr\"{o}dinger equation and all use statistical methods to treat the high-dimensionality of the many-electron
problem efficiently. They make a few well defined approximations that can in principle be systematically removed, albeit at an
exponential cost in general. Diffusion Monte Carlo (DMC)\citep{foulkes_dmc_review} and auxiliary-field quantum Monte Carlo
(AFQMC)\citep{zhang_cpmc,Zhang_phaseless,motta_review} have emerged as the most reliable and general purpose approaches capable of
simulating models\citep{2d_hubbard_benchmark,motta_hydrogen} to ab-initio systems
\citep{williams_transition_metal_simons,zhang_nio,LeeOOMP22020,SaritasRelative2018,DubeckyNoncovalent2016,SantanaCohesive2016,DriverQuantum2010,DevauxElectronic2015,FerlatB2O3VdW2019,CasulaFePairing2013}. Both methods can
be formulated to run efficiently on modern supercomputing architectures\citep{qmcpack,KentQMCPACK2020}.

QMC methods also come with a number of drawbacks. They are expensive relative to DFT or even quantum chemistry approaches for
small to moderately sized systems. They also suffer from finite size effects common to all many-body techniques, which can be slow
to converge. Most seriously, in order to achieve an algorithm that scales only polynomially with system size, both DMC and AFQMC
employ constraints in the Monte Carlo sampling to avoid the Fermion sign problem. This can introduce a significant bias. It is
thus important to assess the quality of the approximations made in both DMC and AFQMC as they become more widely applied in
challenging environments. 

AFQMC and DMC share many similarities: they are projector methods, they use random walkers to sample the many-electron ground
state, and they employ a constraint to control the Fermion sign (phase) problem in DMC (AFQMC). It should be stressed that the
nature of the two constraints is quite different\citep{carlson_no_var}. The methods differ in several additional important ways.
First, DMC has the significant advantage of working in real space and thus in the complete basis set (CBS) limit. AFQMC works in a
finite basis set constructed from plane-waves\citep{suewattana2007phaseless,ma_multiple_proj}, Kohn-Sham
states\citep{ma_dwnf_prl,purwanto_downfolding_jctc}, or a periodized local basis
set\citep{Zhang_phaseless,malone_isdf,motta_kpoint}. Converging the AFQMC results with respect to the single-particle basis set in
solid state calculations introduces a considerable overhead. Second, DMC can incorporate Jastrow factors in the trial wavefunction
to account for electron-electron cusp conditions and capture residual dynamic correlation. Incorporating Jastrow factors in AFQMC
is a challenging prospect\citep{chang_jastrow_afqmc}. Finally DMC, in contrast to AFQMC\citep{carlson_no_var}, can be made fully
variational in the energy, making the assessment and choice of improved trial wavefunctions straightforward in principle. As the
longer established method, DMC has seen by far the widest range of application including to bulk systems with over 1000
atoms\citep{HoodAl2012,LuoTiO22016} and to complex transition metal oxide heterostructures.\citep{SantanaElectron2019} 

Despite the basis-set convergence challenges, AFQMC offers several promising features precisely because it works directly in an
orbital basis. Namely, all-electron, frozen core and non-local pseudopotential calculations can be performed without additional
approximations. In contrast, DMC requires the use of approximations to evaluate non-local potentials, either the original
non-variational locality approximation\citep{MitasLocality} or more recent t-moves methods\cite{CasulaSize2010} that restore the
variational property and increase stability.  The inclusion of spin orbit effects in AFQMC is straightforward requiring very few
algorithmic modifications\citep{rosenberg_so_coup}. Furthermore, many developments from the quantum chemistry community can be
used to improve AFQMC, such as the use of tensor hyper contraction
approaches\citep{thc_1,thc_2,thc_3,lu_isdf,isdf_lin_1,isdf_lin_2,malone_isdf,motta_thc}. Properties other than the total energy
can be more directly accessed\citep{motta_back_prop,motta_itcf_1,motta_itcf_2,vitali_itcf,motta_forces}. Finally, a growing body
of literature suggests that single determinant AFQMC is often more accurate than single determinant DMC
\citep{2d_hubbard_benchmark,motta_hydrogen,williams_transition_metal_simons}. However, little research has been dedicated to the
direct comparison between the two methods in solids and application of multiple determinants.

In this paper we show that both AFQMC and DMC can be made to converge towards the same correlation energy for simple finite size
solids. By employing multi-determinant wavefunctions we show that AFQMC converges more rapidly to the exact ground state energy
than DMC does. DMC on the other hand, shows only a weak basis set dependence. We close by offering some insight into the future
prospects and challenges for the methods.

\emph{Methods.}
Both AFQMC and DMC are projector QMC methods wherein the ground state, $|\Psi_0\rangle$, of the many-electron Hamiltonian,
$\hat{H}$, is determined by
\begin{equation}
|\Psi_0\rangle 
\propto
\lim_{\tau\rightarrow \infty}    
\exp{\left(-\tau \hat{H}\right)} |\Phi_0\rangle
= 
\lim_{\tau\rightarrow \infty} \hat{P}(\tau)|\Phi_0\rangle,
\label{eq:projection}
\end{equation}
where $|\Phi_0\rangle$ is some initial state satisfying $\langle \Psi_0 |\Phi_0\rangle \ne 0$.

In DMC the Schr\"{o}dinger equation is rewritten in imaginary time 
\begin{align}
   {\partial \vert\psi\rangle \over \partial\tau} = -\hat H \vert\psi\rangle,
\end{align}
Where the wavefunction $\vert\psi\rangle$ is expanded over all eigenstates of the Hamiltonian 
\begin{align}
   \vert\psi\rangle=\sum_{i=0} c_{i} \vert\phi_i\rangle
\end{align}
where
\begin{align}
\hat{H}\vert\phi_i\rangle = \epsilon_i \vert\phi_i\rangle 
\end{align}
In real space, any initial state $ \vert\psi\rangle$, that is not orthogonal to the ground state  $\vert\phi _0\rangle$ , will
evolve to the ground state in the long time limit

\begin{align}
\lim_{\tau\to\infty} \psi({\bf R},\tau) = c_0 e^{-\epsilon_{0} \tau}
\phi_{0}({\bf R})
\end{align}
The long limit can be kept finite by introducing an offset $E_T=\epsilon_0$ and the Hamiltonian is separated into the kinetic
energy and potential terms, leading to the diffusion form of the previous equation
\begin{align}
\frac{\partial \psi ({\bf R}, \tau )}{\partial\tau}=
\left[\sum _{i=1}^{N}\frac{1}{2}\nabla _{i} ^{2}\psi({\bf R},\tau)\right]
-(V({\bf R})-E_{T})\psi({\bf R},\tau)
\label{EQ:UnimportanceSample}
\end{align}
Since the potential $V({\bf R})$ is unbounded in Coulombic systems leading to the possible divergence of the rate term $V({\bf
R})-E_{T}$, we use importance sampling for efficiency. We introduce a trial or guiding wavefunction, $\Psi_{G}({\bf R})$,
approximating the ground state wavefunction
\begin{align} f({\bf R},\tau )=\psi_{G}({\bf R})\psi({\bf R},\tau), \end{align}
which is also a solution of the diffusion equation when  $\psi({\bf R},\tau)$ is a solution of the Schr\"{o}dinger equation.

The equation \ref{EQ:UnimportanceSample} becomes:
\begin{eqnarray}
   \frac{\partial f({\bf R},\tau)}{\partial\tau}=
\left[\sum _{i=1}^{N}\frac{1}{2}\nabla _{i} ^{2}f({\bf R},\tau)\right] \nonumber \\
-\nabla \left[\frac{\nabla\psi({\bf R})}{\psi({\bf R})} f({\bf R},\tau) \right]-(E_{L}({\bf R})-E_{T})f({\bf R},\tau)
\end{eqnarray}

$E_{T}$ is a "trial energy" introduced to maintain normalization of the projected solution at large $\tau$ and $E_{L}$ is a "local
energy"  depending on configuration $\{{\bf R}\}$:
\begin{align}E_L({\bf R})=\frac{\hat{H}\psi_{T}({\bf R})}{\psi_{T}({\bf R})} \end{align}

To maintain the fermionic nature of the wavefunction we impose anti-symmetry to the guiding function, also known as the fixed-node approximation\cite{Anderson1976}.  This approximation is variational: the accuracy of DMC depends solely on the quality of the nodes of the trial wavefunction and the fixed-node DMC energy is an upper bound to the exact ground state energy. In order to remove the chemically-inert core electrons, non-local pseudopotentials are introduced and evaluated in DMC using 
T-moves\cite{CasulaSize2010}. 

In contrast, AFQMC is usually formulated as an orbital-space approach in which the
Hamiltonian is written as

\begin{align}
    \hat{H} &= \sum_{ij\sigma} h_{ij} \hat{c}^{\dagger}_{i\sigma}\hat{c}_{j\sigma} + \frac{1}{2}\sum_{ijkl,\sigma,{\sigma}^\prime}v_{ijkl}  
 \hat{c}^{\dagger}_{i\sigma}\hat{c}^{\dagger}_{j\sigma^\prime}\hat{c}_{l\sigma^\prime}\hat{c}_{k\sigma}+E_{II},\label{eq:hamil}\\
            &= \hat{H}_1 + \hat{H}_2 + E_{II},
\end{align}
where $\hat{c}^{\dagger}_{i\sigma}$  and $\hat{c}_{i\sigma'}$ are the fermionic creation and annihilation operators, $h_{ij}$ and $v_{ijkl}$
are the one- and two-electron matrix elements and the constant $E_{II}$ is the ion-ion repulsion energy. The two-body part of the propagator is
then written as an integral over auxiliary fields of one-body propagators using the Hubbard-Stratonovich
transformation\citep{hubbard_strat}. An AFQMC simulation then proceeds by sampling an instance of this propagator and applying it
to a random walker which is defined by a weight and Slater determinant. Unfortunately, for the many-electron Hamiltonian the
propagator will be in general complex, thus giving rise to a phase problem\citep{Zhang_phaseless}.  

To control this phase problem Zhang and Krakauer introduced the phaseless-AFQMC method (ph-AFQMC) \citep{Zhang_phaseless} which
uses an importance sampling transformation and a trial wavefunction to enforce a constraint on the walkers' propagation. With this
approximation ph-AFQMC has been applied to a wide variety of
chemical\citep{Al-Saidi2006,purwanto_cholesky,motta_review,purwanto_chromium_dimer,shee_transition_metals,shee_chem_trans,hao_anions_afqmc}
and solid state\citep{motta_hydrogen,suewattana2007phaseless,zhang_nio,lee_2019_UEG} problems. For problems where static
correlation is important, multi-reference expansions can be employed, such as complete active space self-consistent field
(CASSCF)\citep{purwanto_excited_c2,shee_chem_trans}, selected configuration interaction based
approaches\citep{sharma_dice_1,holmes_dice_2,purwanto_excited_c2}, or non-orthogonal multi-Slater determinant
expansions\citep{scuseria_phf,scuseria_phf_grad,Jimenez_phf,borda_nomsd}.

In this work we attempt to remove the fixed-node and phaseless error in DMC and AFQMC respectively by employing multi-determinant
trial wavefunctions of the form
\begin{equation}
|\Psi_T\rangle = \sum_I c_I |D_I\rangle
\end{equation}
where $|D_I\rangle$ are a set of orthogonal Slater determinants. \\

The expansion is built using CIPSI (configuration interaction using a perturbative selection made iteratively),
a selected CI method introduced a long time ago by Huron {\it et al.}\cite{Huron73} 
In this approach, the CI expansion is built iteratively by selecting at each step some  determinants not present 
in the current variational space based on their estimated contribution to the full CI wave function.
More precisely, denoting $|\Psi_0^{(n)}\rangle$ the CIPSI wavefunction at iteration $n$ (starting, for example, with the Hartree-Fock determinant at $n=0$)
\begin{equation}
|\Psi_0^{(n)}\rangle = \sum_{I} c_I^{(n)} |D_I\rangle
\end{equation}
the perturbative contribution at first-order to the wavefunction of each external determinant $|D_\alpha^{(n)} \rangle$ 
(that is, not belonging to the variational space at this iteration and verifying $\langle D_\alpha^{(n)}|H|\Psi_0^{(n)}\rangle \ne 0 $)
can be quantified using their energy contribution
\begin{equation}
e^{(n)}_\alpha = \frac{ |\langle \Psi_0^{(n)}| {H} | D_\alpha^{(n)} \rangle|^2 }{E^{(n)}_{var} - \langle D_\alpha^{(n)} | {\hat H} | D_\alpha^{(n)} \rangle },
\end{equation}
where $E^{(n)}_{var}$ is the CIPSI variational energy of the wave function at this iteration 
\begin{equation}
E^{(n)}_{var}= \frac{\langle \Psi_0^{(n)}|H|\Psi_0^{(n)}\rangle}{\langle \Psi_0^{(n)}|\Psi_0^{(n)}\rangle}.
\end{equation}
In a first step, a number of external determinants corresponding to the greatest values of $e^{(n)}_\alpha$ are incorporated into the variational space and the Hamiltonian is diagonalized 
to give $|\Psi_0^{(n+1)}\rangle$ and $E^{(n+1)}_{var}$. In practice, the number of 
selected external determinants is chosen so that the size of the variational wave function is roughly doubled at each iteration. 
In the second step, the second-order Epstein-Nesbet energy correction
to the variational energy (denoted as $\EPT^{(n)}$) is computed by summing up the contributions of all external determinants
\begin{equation}
    \EPT^{(n)} = \sum_{\alpha} e^{(n)}_\alpha,
\label{PT2}
\end{equation}
and the total CIPSI energy is given by
\begin{equation}
    E_\text{CIPSI}^{(n)} =  E^{(n)}_{var} + \EPT^{(n)}
\end{equation}
The algorithm is then iterated until some convergence criterion (for example, $|\EPT^{(n)}| \le \epsilon$) is met.
For simplicity, in what follows the superscript $n$ will be dropped from the various quantities.

As the number of selected determinants increases, higher-order perturbational contributions become smaller and the CIPSI 
energy can be used as an estimate of the full CI energy, $E_{FCI}$.
To do that in practice, we have adopted the method recently proposed by Holmes {\it et al.}\cite{holmes_dice_2}
in the context of the semistochastic heat-bath configuration interaction (SHCI) method.
While increasing the number of selected determinants, the CIPSI variational energy, $E_{var}$, is plotted as a function 
of the second-order Epstein-Nesbet energy $\EPT$. For sufficiently large expansions, 
$E_{var} \approx E_{FCI}-\EPT$ and the extrapolated value of $E_{var}$ at $\EPT=0$ is an estimate of the FCI limit. This extrapolation procedure has been shown to be robust,
even for challenging chemical situations.\cite{Holmes_2017,sharma_dice_1,Scemama_2018a, Scemama_2018b, Chien_2018, Garniron_2018, Loos_2018} In what follows, these extrapolated CIPSI results are labeled exFCI.

\emph{Computational details.}
All the QMC calculations were performed with the development version of QMCPACK\citep{qmcpack,KentQMCPACK2020}. PySCF\citep{PYSCF}
was used to run the  DFT simulations and to generate the one- and two-electron integrals within the
B3LYP\cite{B3LYP1,B3LYP2,B3LYP3,B3LYP4} exchange and correlation functional. All calculations were carried out using the
correlation-consistent effective core potentials\cite{Bennett2018,Bennet2017,Wang2019}  and the associated basis sets. We studied
three simple solids in their primitive cells: carbon in the diamond structure (2 atoms per cell), lithium fluoride (2 atoms per cell) and fcc aluminum (4 toms per cell) each at their experimental lattice parameters
of 3.567\AA, 4.0351\AA and 4.046\AA, respectively.

All CIPSI calculations were performed with Quantum Package\cite{QP2}. The iterative process of selection was stopped when the change in $E_{var}+\EPT$ between iteration $n$ and iteration $n+1$ varies with less than $0.5 \times 10^{-4}$ Ha. Total energies of all 3 materials with regards to the basis set size, final number of determinants and value of $\EPT$ are given in the supplemental material\citep{supplement}. The exFCI estimates obtained by extrapolation using $\EPT$ values as explained above are also reported.\citep{supplement}

All DMC calculations used individual Slater-Jastrow trial wavefunctions with one-body, two-body and three-body Jastrow functions.
The total number of determinants used in the trial wavefunction in the DMC runs is explicitly stated in the supplemental material\citep{supplement}. The size of the determinant expansion  corresponds to truncations using the weight of the coefficients ($10^{-4}$, $10^{-5}$,$10^{-6}$ and $10^{-8}$) as a inclusion criterion. The total number of determinants spanned from 15k determinants for the LiF system in its cc-pvDz basis set to 10.5M determinants for Aluminum in its cc-pvQz basis sets. The 50 parameters of the jastrow functions were optimized within variational Monte Carlo (VMC)  with a variant of the linear method of Umrigar and  co-workers~\cite{Umrigar2007} for each system at each determinant truncation and each basis set size. The optimized trial
wavefunction was then used in DMC, using a 0.001 time-step and 32000 walkers. 

The AFQMC simulations used a timestep of $0.005$ Ha$^{-1}$, with a population of 1440 walkers. The pair-branch population control method was
used\citep{wagner_qwalk}. We used the modified-Cholesky decomposition\citep{modified_chol_1,modified_chol_2,modified_chol_3} to
factorize the two-electron integrals and used a convergence threshold of $1\times 10 ^{-5}$. Further simulation details and
convergence studies are in the supplementary material\citep{supplement}.

All input files, output data and scripts necessary to generate the results presented are available at
Ref.~\citenum{materials_data}. Note that neither the determinant coefficients nor the orbitals were reoptimized after the
initial DFT and CIPSI procedure for either DMC or AFQMC, which could accelerate convergence.

\begin{figure}
\includegraphics{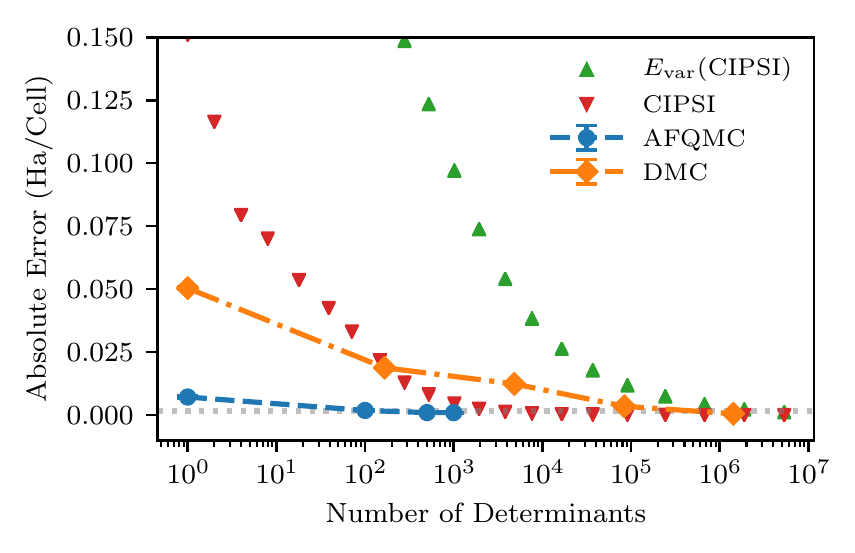}
\caption{Convergence of the phaseless AFQMC and fixed-node DMC error in the total energy for diamond structure Carbon in the cc-pVTZ basis
set with the size of the multi-determinant expansion. For AFQMC the error was computed relative to CIPSI total energy in the same
basis set. For DMC the error was computed with respect to the DMC result with $N_D=1\times 10^7$\label{fig:det_conv}. Horizontal
dashed line represents chemical accuracy of 1.6 mHa/Cell. $E_{var}(\mathrm{CIPSI})$ is the variational energy of the CIPSI wavefunction,
while the full CIPSI energy includes the second order perturbation theory correction.
}
\end{figure}

\begin{figure*}
\includegraphics[width=\textwidth]{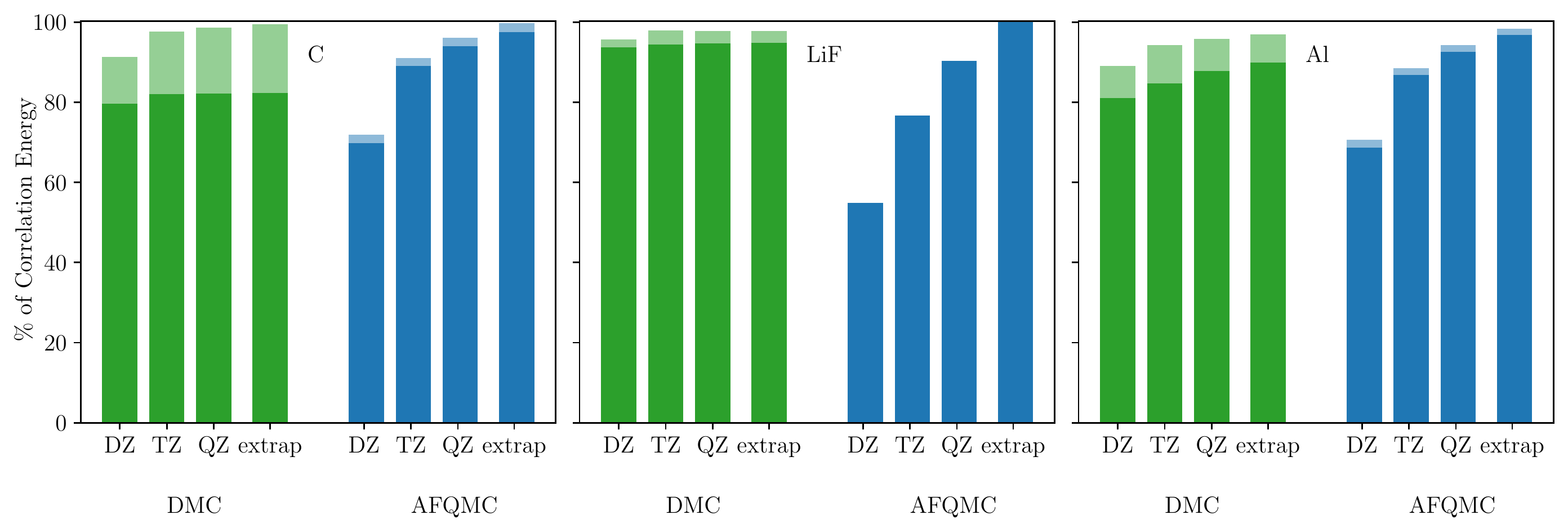}
\caption{Comparison between the amount of correlation energy captured by DMC and AFQMC with single-determinant (dark bars) and multi-determinant (light bars) trial wavefunctions for unit cells of Carbon (Diamond), LiF and Al. For both DMC and AFQMC we compare to the converged CBS-limit exFCI correlation energies computed with CIPSI.~\label{fig:final_comp}}
\end{figure*}

\emph{Results.}
In \cref{fig:det_conv} we show the results for Carbon (Diamond) in the cc-pVXZ basis sets, where X is the cardinality of the basis
set. Both AFQMC and DMC energies converge faster than the CIPSI variational energy.
The phaseless error of AFQMC converges faster compared to the fixed-node error in DMC.
 We see that the ph-AFQMC results is
within chemical accuracy of the converged total energy using approximately 100 determinants. In contrast the CIPSI variational
energy required $\mathcal{O}(10^7)$ determinants to reach this level of accuracy and $\mathcal{O}(10^4)$ determinants for the CIPSI energy. 
Note that reducing the number of determinants by three orders of magnitude when passing from the variational to the full CIPSI energy 
illustrates how much the second-order energy correction $\EPT$ is efficient at enhancing the convergence.
We see that DMC also converges systematically with the size of the multi-determinant expansion, although it requires $\mathcal{O}(10^6)$ determinants
to reach the same level of accuracy as AFQMC.
We observed the same trends observed in carbon as in aluminum and lithium fluoride. \cref{fig:final_comp} presents the fraction of correlation energy captured by each method relative to the estimated exact correlation energy for each material.
The exact correlation energies can be obtained here with CIPSI 
since the regime where energies are converged 
both as a function of the number of determinants for a given basis set (extrapolation to  $\EPT=0$ to get exFCI estimates as explained above) and of the basis set (CBS limit) 
can be attained for these simple systems.
We estimate the basis set extrapolation errors to be 
of the order of 1-2 mHa/Cell.
Details are given in the supplemental material.
We define the percentage of correlation energy recovered as  
\begin{equation}
\text{\% of Correlation Energy} = 100 \times \left|\frac{E_c(\text{QMC})}{E_c(\text{exact})}\right|,
\end{equation}
where the correlation energy is defined as $E_c = E-E_{\text{HF}}$ where $E_{\text{HF}}$ is the restricted Hartree--Fock total
energy in the CBS limit. 

We show it is possible to converge both AFQMC and DMC towards the same correlation energy, once systematic biases are removed. We estimate the basis set extrapolation for the exFCI values could introduce errors
on the order of 1-2 mHa/Cell. These are most pronounced in case of LiF and Al. Thus, AFQMC and DMC may agree
better if larger, augmented, or more optimally chosen basis sets were used, since less extrapolation would be required.

We see that the relative gain in correlation energy by using a modest multiple determinant trial in AFQMC
is small (on the order of a few percent), whilst the largest error from the CBS-limit exFCI results is, unsurprisingly, the basis set
error. Notably we see for LiF the phaseless error is essentially zero, and is largest in the 4-atom cell of Al. The DMC results in
contrast show a much larger dependence on trial wavefunction, with the single determinant correlation energies exhibiting up to a
20\% error. In the smallest tested basis (double zeta), DMC is always more
accurate for total energies than AFQMC performed in the same basis. However the DMC improves slowly with multiple determinants and with increased
basis set size, while the accuracy of AFQMC gains rapidly. \cref{tab:conv} summarizes our results for single determinants and for
the largest number of determinants that were run for each material.

For the case of aluminum we begin to see the ultimate limitation of all methods to systematically remove their respective
constraint error. Obtaining reliable estimate of exFCI correlation energies was challenging, due to the difficulty in reaching the linear regime, $E_{var} \approx E_{FCI}-\EPT$,
required for the extrapolation step\citep{supplement}. 

In the case of AFQMC, we found it challenging to reach the same
level of convergence in the cc-pVQZ basis set as in the cc-pVDZ basis set, due to the computational cost which grows with the basis set size. Thus, the CBS extrapolated AFQMC value in the case is not fully reliable.  Ultimately, we begin to see
the limitation of the present multi-determinant trial wavefunctions. This is perhaps unsurprising as we
would expect that the amount of correlation energy captured for a fixed multi-determinant size should decay exponentially with
system size. Put another way, this is a manifestation of the lack of size consistency (extensivity) of truncated multi-determinant
expansions. Nevertheless, for these relatively simple cases, we managed to obtain reliably converged total energies accurate to
roughly 1-2 mHa/Cell, which should serve as useful benchmarks for the future studies.

\begin{table}[h!]
\begin{tabular}{llll}
\toprule
  &    C & LiF & Al \\ 
\hline
  $E_\mathrm{HF}$ & -10.2381 & -31.5559& -7.7987\\ 
  $E$ & -10.5569 & -31.9038 & -8.2158  \\
  $E_c$ & \ -0.3187 & \ -0.3479 & -0.4171 \\
  \hline
  AFQMC(SD) & \ -0.0077(9) & \ \ 0.0008(5) & -0.0130(8) \\
  AFQMC(MD) & \ -0.0007(3) & \ -0.0008(7) & -0.0067(6) \\ 
  \hline
  DMC(SD) & \ -0.0563(8) & \ -0.0180(8)& -0.042(1)\\
  DMC(MD) & \ -0.002(2) &  \ -0.008(2) & -0.0129(8) \\ 
   \hline
\end{tabular}
\caption{Converged CIPSI total ($E$), Hartree-Fock ($E_{\mathrm{HF}}$), and correlation $E_c$ energies for the systems considered here. Also presented is the error in the basis set extrapolated AFQMC and DMC correlation energies. Energies are in Hartree/Cell. SD indicates single determinant results while MD indicates results from the largest multi determinant trial wavefunctions for each system.\label{tab:conv}}
\end{table}

\emph{Discussion and Conclusions.}
We have shown it possible to systematically converge CIPSI, AFQMC and DMC to the exact ground state total energy of three simple
finite size solids. We have shown that the phaseless constraint in AFQMC is often much smaller than the fixed-node error in DMC, an
observation that has not been quantified before in solids. We also showed that the phaseless error can be removed by using smaller
multi-determinant expansions than DMC. At the same time we found that AFQMC exhibits a much larger basis set error than DMC. 

In light of these findings it is clear that the most important issues for the application of AFQMC in solids is the development of
robust basis set correction techniques to accelerate convergence and the development of optimized basis sets. For DMC it is the
need to develop more accurate compact trial wavefunctions that converge similarly efficiently as in AFQMC. This could be via
optimized orbitals and improved multiple determinant selection schemes, a full reoptimization of determinant coefficients, or
wholly different wavefunctions such as Geminals, Pfaffians, or backflow. For practical applications where  relative energies
rather than total energies are used, convergence of both methods is likely to be better due to cancellation of errors. Indeed
cohesive energies computed from single-determinant DMC are often very
accurate\cite{SaritasRelative2018,DubeckyNoncovalent2016,SantanaCohesive2016,DriverQuantum2010}. Further work should investigate
why single determinant DMC errors are so large by, for example, investigating the magnitude of the locality error.

Looking to the future, an important topic not addressed is the treatment of finite size effects. In light of our findings, it
seems highly unlikely that highly converged multi-determinant trial wavefunctions could be used to obtain thermodynamic limit
total energies in QMC. Nevertheless, it may be possible to obtain corrections using simpler wavefunctions, and apply this
correction to more accurate small unit-cell results. Interestingly, we found that the phaseless error in AFQMC is roughly
independent of the basis set size\citep{supplement}. Further work should include the investigation of the effect of basis set and
orbital optimizations, and the convergence of properties other than the total energy with respect to the trial wavefunction.
Ultimately, we hope that our results will serve as helpful reference and motivate the development of compact and efficient trial wavefunctions for both AFQMC and DMC. 

\emph{Acknowledgement.}
Notice: This manuscript has been authored by UT-Battelle, LLC, under Contract No. DE-AC0500OR22725 with the U.S. Department of
Energy. The United States Government retains and the publisher, by accepting the article for publication, acknowledges that the
United States Government retains a non-exclusive, paid-up, irrevocable, world-wide license to publish or reproduce the published
form of this manuscript, or allow others to do so, for the United States Government purposes. The Department of Energy will
provide public access to these results of federally sponsored research in accordance with the DOE Public Access Plan
(\url{http://energy.gov/downloads/doe-public-access-plan}). This work was supported by the U.S. Department of Energy (DOE), Office
of Science, Basic Energy Sciences, Materials Sciences and Engineering Division, as part of the Computational Materials Sciences
Program and Center for Predictive Simulation of Functional Materials (CPSFM). MC was supported by the ANR PhemSpec project, grant ANR-18-CE30-0025-02 of the French Agence Nationale de la Recherche
and AB and MC were partially supported by the international exchange program CNRS-PICS, France-USA. An award of computer time was provided by the
Innovative and Novel Computational Impact on Theory and Experiment (INCITE) program. All CIPSI and DMC calculations used resources
of the Argonne Leadership Computing Facility, which is a DOE Office of Science User Facility supported under Contract
DE-AC02-06CH11357. AFQMC calculations received computing support from the LLNL Institutional Computing Grand Challenge program.
The work of FDM and MAM was performed under the auspices of the U.S. DOE by LLNL under Contract No. DE-AC52-07NA27344. Sandia
National Laboratories is a multi-mission laboratory managed and operated by National Technology \& Engineering Solutions of
Sandia, LLC, a wholly owned subsidiary of Honeywell International Inc., for the U.S. Department of Energy’s National Nuclear
Security Administration under contract DE-NA0003525.
\bibliography{refs}

\begin{thebibliography}{89}
\bibitem [{mat()}]{materials_data}%
  \BibitemOpen
  \href@noop {} {}\bibinfo {howpublished} {All simulation data is available at
  [URL to be inserted].}\BibitemShut {Stop}%
\end{thebibliography}
\end{document}